\documentclass{article}
\usepackage[utf8]{inputenc}
\usepackage{graphicx}
\usepackage{blindtext}
\usepackage{amsmath}
\usepackage{amssymb}
\usepackage{booktabs} 
\usepackage{amsthm} 
\usepackage{bbold}
\usepackage{amsfonts, bbm, paralist}
\usepackage{geometry} 
\usepackage{enumitem}
\usepackage{array}
\usepackage{multirow}
\usepackage{algorithm}
\usepackage{algpseudocode}
\usepackage{hyperref}
\hypersetup{
    hidelinks 
}
\newcommand{\pkg}[1]{\textbf{#1}}
\newcommand{\CRANpkg}[1]{\href{https://CRAN.R-project.org/package=#1}{\pkg{#1}}}

\makeatletter
\DeclareRobustCommand\code[1]{%
  \ifmmode
    \mathtt{#1}
  \else
    \leavevmode\unskip\texttt{#1}\ignorespaces
  \fi
}
\makeatother
\usepackage[backend=biber, style=apa, natbib=true]{biblatex}
\title{\textbf{ProfileGLMM: a R Package Extending Bayesian Profile Regression using Generalised Linear Mixed Models.}}
\author{}
\author{Matteo Amestoy,
 Mark A. van de Wiel,
 Wessel N. van Wieringen}

\begin{document}
\maketitle
\paragraph{ABSTRACT}
ProfileGLMM is an R package integrating Generalised Linear Mixed Models (GLMMs) as the outcome model for Bayesian profile regression. This statistical framework simultaneously i) explains the variation in the outcome and ii) clusters the observations based on a specified set of interdependent clustering covariates. The derived cluster memberships are then incorporated, alongside others, as explanatory variables in the regression to model the outcome. This framework efficiently handles complex, highly correlated covariate structures whose direct inclusion in a standard regression model would be statistically sub-optimal. ProfileGLMM significantly extends Bayesian profile regression's scope by resolving two key constraints of previous implementations: 1) it allows the analysis of hierarchical and longitudinal data structures through the inclusion of random effects, and 2) it enables the study of interactions between latent clusters and other observable covariates. ProfileGLMM accommodates various data types, supporting both continuous or binary outcomes and both categorical and continuous clustering covariates. Built on fast Rcpp code with minimal mandatory parameters, ProfileGLMM offers a flexible analytical tool. It significantly enhances the utility of profile regression for researchers in fields such as epidemiology, social sciences, and clinical studies dealing with complex data.
\clearpage

\section{Introduction}\label{introduction}

This paper introduces the \CRANpkg{ProfileGLMM} package, an R implementation freely available on The Comprehensive R Archive Network (CRAN). The \CRANpkg{ProfileGLMM} package builds on the method originally presented in \citet{amestoy_bayesian_2025} and extends the functionality of Bayesian profile regression by integrating Generalised Linear Mixed Models (GLMMs) as the outcome model.

Bayesian profile regression, initially formulated by \citet{molitor_bayesian_2010}, is a powerful statistical framework designed to model an outcome variable (\(\mathbf{y}\)), by a set of
explanatory covariates that decomposes into two distinct groups. The first group, regression covariates (\(\mathbf{x}\)), consists of conventional explanatory variables (e.g., confounding individual characteristics).
The second group, referred to as clustering covariates (\(\mathbf{u}\)), typically comprises variables characterized by strong interdependence (e.g., highly correlated environmental exposures \citet{amestoy_bayesian_2025} or detailed questionnaire responses \citet{molitor_bayesian_2010}). Due to these interdependencies, direct inclusion of \(\mathbf{u}\) into a regression model is often statistically sub-optimal. To address the complexity of \(\mathbf{u}\), profile regression non-parametrically clusters the observations based on these covariates. Consequently, instead of using \(\mathbf{u}\) in the regression model, the resulting cluster memberships are then incorporated as indicator variables alongside \(\mathbf{x}\). Crucially, profile regression provides a unified and efficient inference strategy by estimating the parameters for both the clustering and regression components simultaneously, unlike traditional sequential approaches.

The standard Bayesian profile regression methodology has been primarily applied to outcomes derived from cross-sectional data \citep{molitor_bayesian_2010, papathomas_exploring_2012}, as demonstrated by the \CRANpkg{PReMiuM} R package \citep{liverani_premium_2015}.
This implementation offers flexibility across various outcome types (including categorical, count, and continuous data) and has been recently extended with longitudinal continuous data \citep{rouanet_bayesian_2023}. However, the approach proposed by \citet{rouanet_bayesian_2023} is limited to the clustering of individuals and lacks the necessary flexibility to effectively model general hierarchical data structures. Moreover, it does not account for potential interactions between the latent clusters and observable covariates (see \citet{amestoy_bayesian_2025} for discussion on this necessity).

\CRANpkg{ProfileGLMM} directly addresses this critical gap by embedding GLMMs within the Bayesian profile regression framework. This integration significantly increases the scope and utility of Bayesian profile regression, allowing for the analysis of outcomes derived from complex hierarchical and repeated-measures designs, such as longitudinal studies or clustered observations. Specifically, the inclusion of random effects enables the model to account for unobserved heterogeneity and within-subject correlation, which are mandatory for valid inference in these complex data structures.

The remainder of this paper is structured as follows. In \hyperref[sec:math]{Section 2}, we provide a formal specification of the statistical models underlying the \CRANpkg{ProfileGLMM} package. \hyperref[sec:example]{Section 3} is dedicated to describing the functionality and application of the \CRANpkg{ProfileGLMM} package through illustrative examples.
Specifically, \hyperref[sec:expoLMM]{Subsection 3.1} utilizes a simulated exposure dataset to demonstrate the default sequential usage of the package functions, outlining a typical analysis workflow. Following this, \hyperref[sec:piecewise]{Subsection 3.2} showcases the versatility of \CRANpkg{ProfileGLMM} by applying it to perform a piecewise linear fit, highlighting its capacity for flexible regression modeling. Finally, in \hyperref[sec:advanced]{Section 4}, we detail the tuning of optional parameters and present practical considerations pertaining to the robust and efficient usage of the primary functions.

\section{Bayesian profile regression of a generalised linear mixed model}\label{sec:math}

\subsection{Bayesian profile regression}\label{bayesian-profile-regression}

Bayesian profile regression considers a dataset comprising \(n\) observations of the form \(\mathcal{D} = \{(y_i, \mathbf{u}_i,\mathbf{x}_i) \mid i = 1, ..., n\}\). Each observation \(i\) consists of an outcome variable, \(y_i\), a corresponding vector of clustering covariates, \(\mathbf{u}_i\) and a vector of regression covariates \(\mathbf{x}_i\). Bayesian profile regression models the joint distribution of \((y_i, \mathbf{u}_i)\) conditionally on \(\mathbf{x}_i\) using a mixture model. The density function of this model is expressed as:

\begin{align}
p(y_i,\mathbf{u}_i|\boldsymbol{\pi},\boldsymbol{\theta},\mathbf{x}_i,\boldsymbol{\alpha}) &= \sum_{c=1}^{C}\pi_c f(y_i, \mathbf{u}_i | \boldsymbol{\theta}_c,\mathbf{x}_i,\boldsymbol{\alpha}) \\ 
&= \sum_{c=1}^{C}\pi_c f_y(y_i | \boldsymbol{\theta}_c^y,\mathbf{x}_i,\mathbf{u}_i,\boldsymbol{\alpha}) f_u(\mathbf{u}_i | \boldsymbol{\theta}_c^u),
\end{align}

\sloppy where \(f(.|\boldsymbol{\theta}_c,\mathbf{x}_i,\boldsymbol{\alpha})\) is the mixture component density, decomposable into \(f_y(y_i | \boldsymbol{\theta}_c^y,\mathbf{x}_i,\mathbf{u}_i,\boldsymbol{\alpha})\) representing the outcome or regression distribution and \(f_u(\mathbf{u}_i | \boldsymbol{\theta}_c^u)\) representing the assignment distribution. The vector \(\boldsymbol{\alpha}\) contains the population constant parameters shared by all the observations and \(\boldsymbol{\pi} = \{\pi_c\}\) represents the mixture weights. Consistent with the model's original formulation, the number of mixtures, \(C\), is treated as infinite. The computational challenge of this infinite state space is addressed in detail in \hyperref[sec:Inference]{Subsection 2.4}.

In standard Bayesian profile regression, the outcome \(y_i\) and the clustering covariates \(\mathbf{u}_i\) are assumed to be independent conditional on the cluster-specific parameters (i.e., \(f_y(y_i | \boldsymbol{\theta}_c^y,\mathbf{x}_i,\mathbf{u}_i,\boldsymbol{\alpha}) = f_y(y_i | \boldsymbol{\theta}_c^y,\mathbf{x}_i,\boldsymbol{\alpha})\)). Our proposed profile-GLMM framework relaxes this conventional constraint, allowing the outcome distribution to also depend on the clustering covariates \(\mathbf{u}_i\). In \hyperref[sec:piecewise]{Subsection 3.2}, we demonstrate how this feature can be leveraged to construct an estimation method for a piecewise linear function.

In the next subsections, we will detail the types of outcome and assignment distributions that our package can accommodate.

\subsection{Assignment distribution}\label{assignment-distribution}

We now outline the assignment distribution, \(f_u(\mathbf{u}_i | \boldsymbol{\theta}_c^u)\), following the structure proposed by \citet{liverani_premium_2015}. The model is designed to accommodate clustering covariates \(\mathbf{u}_i\) that are either continuous or categorical, or a mixture of both.\\
For purely categorical covariates, \(\mathbf{u}_i= (u_{i,1},\cdots, u_{i,q^u})\), each of the \(q^u\) covariates is assumed to follow an independent multinomial distribution, \(u_{i,j}|\boldsymbol{\theta}_c^u\sim\mathcal{M}(\boldsymbol{\phi}_{c,j})\). The vector \(\boldsymbol{\phi}_{c,j}\) represents the probabilities associated with the categories of \(u_{i,j}\) for cluster \(c\). For purely continuous covariates, a multivariate Gaussian distribution is employed: \(\mathbf{u}_i|\boldsymbol{\theta}_c^u \sim \mathcal{N}(\boldsymbol{\mu}_c,\boldsymbol{\Sigma}_c)\). In the case where the covariate vector \(\mathbf{u}_i\) comprises a mixture of continuous variables and \(\ell\) categorical variables, we assume independence between these two types. Accordingly, the assignment parameters are partitioned as \(\boldsymbol{\theta}_c^u = \{\boldsymbol{\phi}_{c,1},\cdots,\boldsymbol{\phi}_{c,\ell},\boldsymbol{\mu}_c,\boldsymbol{\Sigma}_c\}\).

\subsection{Generalised linear mixed model as outcome regression}\label{generalised-linear-mixed-model-as-outcome-regression}

The regression distribution, denoted as \(f_y(y_i | \boldsymbol{\theta}_c^y,\mathbf{x}_i,\mathbf{u}_i,\boldsymbol{\alpha})\), is the novel element of our method. Our package employs GLMMs as the outcome framework. GLMMs provide a more comprehensive structure, addressing the hierarchical nature and intrinsic dependencies found in clustered or repeated-measures datasets. Currently, our package includes implementations for linear mixed models for continuous outcomes and probit mixed models for binary outcomes.

We first outline the specifics of the linear mixed model scenario and then transition to the probit framework. We begin by considering, without any loss of generality, that the hierarchical nature of the data arises from repeated measurements gathered from a group of \(m\) individuals. To simplify the model's notation, we establish a mapping function \(g: [1,n] \rightarrow [1,m]\) which connects each observation index \(i \in [1,n]\) with the corresponding individual index, thus associating each observation with its originating individual. For an observation \(i\) that is part of latent cluster \(c\), we represent the continuous outcome variable \(y_i\) in the following manner:

\begin{align}
    y_i &= \mathbf{x}^{\text{Fe}}_i\boldsymbol{\beta} + \mathbf{x}^{\text{Re}}_i\boldsymbol{\eta}_{g(i)}+\mathbf{x}^{\text{Int}}_i\boldsymbol{\gamma}_c+\epsilon_i, \label{eq:LMM}\\
    \boldsymbol{\eta}_{j}&\sim\mathcal{N}(0,\mathbf{W}^{\text{Re}}),\,\,\, \text{for all }j \in [1,m]\nonumber \\ 
    \epsilon_i &\sim\mathcal{N}(0,\sigma^2)\nonumber
\end{align}

As is common with mixed models, the vectors \(\mathbf{x}^{\text{Fe}}_i\) and \(\mathbf{x}^{\text{Re}}_i\) represent specific subsets of the covariates \(\mathbf{x}_i\cup\mathbf{u}_i\), corresponding to fixed effect covariates and random effect covariates, respectively. When \(\mathbf{x}^{\text{Re}}_i\) is a constant, the model is known as a random intercept model. Conversely, if \(\mathbf{x}^{\text{Fe}}_i=\mathbf{x}^{\text{Re}}_i\), it indicates that all covariates have random slopes. The vector \(\mathbf{x}^{\text{Int}}_i\) is another subset of covariates \(\mathbf{x}_i\cup\mathbf{u}_i\), which are hypothesized to interact with the clustering covariates \(\mathbf{u}_i\). If \(\mathbf{x}^{\text{Int}}_i\) is limited to a constant intercept, then the effect of the exposures on the outcome is confined to a specific profile mixture offset. In essence, \(\mathbf{x}^{\text{FE}}_i,\mathbf{x}^{\text{RE}}_i,\mathbf{x}^{\text{Int}}_i\) are all (potentially overlapping) subsets of the regression covariates \(\mathbf{x}_i\cup\mathbf{u}_i\), representing the fixed effects, the individual random effects, and the variables interacting with latent exposure clusters, respectively.\\
The vector \(\boldsymbol{\beta}\) signifies the fixed effect coefficients, which remain constant across all observations, while \(\boldsymbol{\eta}_{g(i)}\) denotes the random effect parameters common to all observations of the subject \(g(i)\). These random effects are modeled to follow a Gaussian distribution with a constant random effect covariance matrix \(\mathbf{W}^{\text{Re}}\). The vector \(\boldsymbol{\gamma}_c\) is a profile mixture-specific parameter that applies to all observations within cluster \(c\). The error term \(\epsilon_i\) is homoscedastic with a variance of \(\sigma^2\). Using the notation from the previous section, the population parameters are identified as \(\boldsymbol{\alpha} = \{\boldsymbol{\beta},\sigma^2,\mathbf{W}^{\text{Re}}\}\), and the mixture parameters are expressed as \(\boldsymbol{\theta}_c^y = \boldsymbol{\gamma}_c\).

For the probit model, the underlying structure is defined by incorporating a continuous latent variable, denoted \(y_i^{\ast}\), such that the observed binary outcome \(y_i\) is determined by whether \(y_i^{\ast}\) exceeds zero. This relationship is formally specified through the following set of equations:
\begin{align}
y_i^\ast &= \mathbf{x}^{\text{Fe}}_i\boldsymbol{\beta} + \mathbf{x}^{\text{Re}}_i\boldsymbol{\eta}_{g(i)}+\mathbf{x}^{\text{Int}}_i\boldsymbol{\gamma}_c+\epsilon_i \\
y_i &= \mathbb{1}_{y_i^\ast>0}\nonumber \\
    \boldsymbol{\eta}_{j}&\sim\mathcal{N}(0,\mathbf{W}^{\text{Re}}),\,\,\, \text{for all }j \in [1,m]\nonumber  \\
    \epsilon_i &\sim\mathcal{N}(0,1)\nonumber.
\end{align}
The \(\mathbb{1}\) function represents the indicator function, which returns a value of \(1\) when the specified condition is satisfied and \(0\) otherwise. In the context of the probit mixed model, the population parameters are defined as \(\boldsymbol{\alpha} = \{\boldsymbol{\beta}, \mathbf{W}^{\text{Re}}\}\), while the mixture parameters are denoted by \(\boldsymbol{\theta}_c^y = \boldsymbol{\gamma}_c\).
This latent variable formulation is equivalent to stating the conditional probability of the outcome as:
\(p(y_i=1) = \boldsymbol{\Phi}(\mathbf{x}^{\text{Fe}}_i\boldsymbol{\beta} + \mathbf{x}^{\text{Re}}_i\boldsymbol{\eta}_{g(i)}+\mathbf{x}^{\text{Int}}_i\boldsymbol{\gamma}_c)\)
where \(\boldsymbol{\Phi}(\cdot)\) is the cumulative distribution function (CDF) of the standard normal distribution.

\subsection{Parameters and priors}\label{sec:Inference}

In the following section, we will detail our selected prior distributions for the mixture components, then outline those for the assignment distribution, and finally, specify those for the outcome distribution.

As is customary with infinite mixture models, a Dirichlet process prior is specified. Following \citet{liverani_premium_2015}, a stick breaking process prior \citep{ishwaran_gibbs_2001, pitman_combinatorial_2006} is imposed on the mixture weights \(\pi_c\):
\begin{align*}
    V_c&\sim \beta(1,\zeta),\, \text{ i.i.d for } c\in\mathbb{Z}^+\\
    \pi_1 &= V_1,\,\,\,\,\pi_c = V_c\prod_{j=1}^{c-1}(1-V_j)
\end{align*}
Here, \(\beta(1,\zeta)\) denotes the beta distribution with shape parameters 1 and \(\zeta\). The concentration parameter \(\zeta\), which governs the expected number of mixture components, is subsequently estimated within the model using a Gamma prior.

To enhance computational tractability, we adopt the truncated Dirichlet process mixture model (DPMM) framework, as proposed by \citet{molitor_bayesian_2010}. This approach assumes a finite, sufficient maximum number of clusters, effectively restricting the latent parameter space to a manageable dimension while remaining virtually indistinguishable from an infinite-dimensional prior \citep{ishwaran_gibbs_2001}. Compared to standard profile regression, this truncation eliminates the need for the computationally demanding Metropolis-within-Gibbs step often required for sampling cluster allocations.

We now define the prior distributions for the cluster assignment distribution. For every categorical covariate \(u_{i,j}\) within \(\mathbf{u}_{i}\), a conjugate Dirichlet prior is used for \(\boldsymbol{\phi}_{c,j}\). For the continuous covariates that make up \(\mathbf{u}_{i}\), a conjugate normal-inverse-Wishart prior is applied to \((\boldsymbol{\mu}_c,\boldsymbol{\Sigma}_c)\). When the covariate vector \(\mathbf{u}_i\) includes both continuous and categorical variables, the respective parameters are given independent priors.

In terms of outcome distribution, we utilize the conventional conjugate framework. We apply a normal-gamma prior to the fixed effect parameter and error variance term \((\boldsymbol{\beta},\sigma^2)\), while an inverse-Wishart prior is designated for the variance matrix of the random effects \(\mathbf{W}^{\text{Re}}\). As for the regression parameters of the latent clusters \(\gamma_c\), a Gaussian prior with covariance \(\mathbf{W^{\text{Int}}}\) is used. The impact of the latent factor matrix, \(\mathbf{W^{\text{Int}}}\), tends to be significant and is typically not known in advance. To manage this, we estimate \(\mathbf{W^{\text{Int}}}\) by integrating it into the model, using an inverse-Wishart prior.

\subsection{Posterior sampling with a Gibbs sampler}\label{posterior-sampling-with-a-gibbs-sampler}

We employ a Markov Chain Monte Carlo (MCMC) method to sample from the posterior distribution of the model parameters. These parameters are categorized as follows:

\begin{itemize}
\item
  outcome distribution parameters: \(\{\boldsymbol{\alpha},\boldsymbol{\theta}_1^y,\dots,\boldsymbol{\theta}^y_C,\boldsymbol{W}^\text{Int}\}\), comprising population constant parameters \(\boldsymbol{\alpha} = \{\boldsymbol{\beta},\sigma^2,\boldsymbol{W}^\text{Re}\}\), mixture-dependent parameters \(\boldsymbol{\theta}^y_c = \boldsymbol{\gamma}_c\) and \(\boldsymbol{W}^\text{Int}\) the parameters of the mixture dependent parameters' prior.
\item
  asssignment distribution mixture dependent parameters: \(\boldsymbol{\theta}^u = \{\boldsymbol{\theta}_1^u,\dots,\boldsymbol{\theta}^u_C\}\), where \(\boldsymbol{\theta}_c^u = \{\boldsymbol{\phi}_{c,1},\cdots,\boldsymbol{\phi}_{c,l},\boldsymbol{\mu}_c,\boldsymbol{\Sigma}_c\}\),
\item
  mixture parameters: \(\boldsymbol{\theta}^m = \{\boldsymbol{\pi},\zeta\}\).
\end{itemize}

As is common practice in mixture model inference \citep{liverani_premium_2015}, we introduce a vector of latent allocation variables, \(\mathbf{Z}\). For each observation \(i\), the latent variable \(\mathbf{Z}_i\) dictates the mixture component associated with that observation, such that:
\[p(y_i,\mathbf{u}_i|\boldsymbol{\theta},\mathbf{Z}_i=c) = f(y_i,\mathbf{u}_i|\boldsymbol{\theta}_c)\text{ and } p(\mathbf{Z}_i=c|\boldsymbol{\pi})=\pi_c.\]
Essentially, \(\mathbf{Z}_i\) represents a cluster membership, where all observations sharing the same \(\mathbf{Z}_i\) value are assigned to the same mixture component and consequently share the same underlying distribution parameters.

By incorporating \(\mathbf{Z}\) into the parameter space, the posterior distributions of the outcome and assignment parameters become conditionally independent given \(\mathbf{Z}\). Since each block represents a relatively simple conditional model, we can derive an efficient Gibbs sampler.

The design of our Gibbs sampler is inspired by \citet{ishwaran_gibbs_2001} (for the Dirichlet process mixture model sampling) and \citet{liverani_premium_2015} (for profile regression model inference). A detailed step-by-step description of the algorithm is presented in the Appendix.

\subsection{Inference and post-processing of the posterior samples}\label{postPro}

For population-level constant parameters, such as the fixed effect coefficients \(\boldsymbol{\beta}\), standard Bayesian posterior distribution analysis can be readily conducted. However, extracting information from the posterior samples for latent parameters is challenging due to the label switching problem in MCMC samples. Each sample has valid cluster assignments, but inconsistent labeling prevents direct parameter aggregation. To solve this, we use the method from \citet{molitor_bayesian_2010}, \citet{liverani_premium_2015}, creating a representative clustering via an average dissimilarity matrix \(\mathbf{S}\), which shows how often pairs of observations cluster together. We then use this matrix to identify a representative clustering \(\mathbf{Z}^{\ast}\).

Once the representative clustering is obtained, we can estimate the cluster-specific parameters and their associated uncertainties by integrating over the entire MCMC chain, effectively leveraging the information from all the sampled latent configurations.
We demonstrate the methodology using the outcome mixture specific parameter \(\boldsymbol{\gamma}\), recognizing that the same procedure is applicable to the cluster-specific assignment distribution mixture \(\boldsymbol{\theta}^u\). For each cluster \(c\) identified by the representative clustering \(\mathbf{Z}^{\ast}\), we compute the mean a posteriori estimate, \(\boldsymbol{\gamma}^\ast_c\), of the parameter by aggregating the posterior samples across both MCMC iterations and all the observations assigned to cluster \(c\). The formula for this mean estimate is: \[\boldsymbol{\gamma}^\ast_c = \frac{1}{HN_c}\sum_{h=1}^H\sum_{i|\mathbf{Z}^{\ast}_i=c}\boldsymbol{\gamma}^{(h)}_{\mathbf{Z}^{(h)}_i}.\]
Here, \(\boldsymbol{\gamma}^{(h)}_{\mathbf{Z}^{(h)}_i}\) denotes the sampled parameter for the \(h^{\text{th}}\) MCMC sample associated with observation \(i\). \(H\) represents the total number of samples retained in the chain, and \(N_c\) is the number of observations such that \(\mathbf{Z}^{\ast}_i=c\). In effect, the posterior sample pool for the parameter of a given cluster \(c\) is formed by aggregating all parameters \(\boldsymbol{\gamma}^{(h)}_{i}\) across all \(H\) MCMC samples and all \(N_c\) observations for which the corresponding observation \(i\) is assigned to that cluster. Moreover, credible intervals and other standard Bayesian statistical measures can be computed using analogous methodologies.

\section{Functionalities and core components illustrated}\label{sec:example}

The Bayesian Profile-GLMM methodology has been implemented in the R package \CRANpkg{ProfileGLMM} \citep{amestoy_profileglmm_2025}, which is freely available on CRAN.
The core structure of the package is organized around a sequence of functions designed to facilitate model building, MCMC sampling, and posterior inference. An overview of these primary functions is provided in Table \ref{tab:Table1}.

\begin{table}[h!]
\centering
\begin{tabular}{| >{\raggedright\arraybackslash}m{0.31\textwidth} | >{\raggedright\arraybackslash}m{0.28\textwidth} | >{\raggedright\arraybackslash}m{0.36\textwidth} |}
 \hline
 \textbf{Function} & \textbf{Description} & \textbf{Mandatory Input} \\
 \hline
  \small{\code{profileGLMM\_} \code{preprocess}} & Builds the model and initializes all the components. & \begin{itemize}[leftmargin=6pt, itemsep=1pt]
  \item Regression (\code{'linear'} or \code{'probit'})
  \item Covariate structure
  \item Data
  \item Maximal number of clusters
  \item Presence of intercepts
 \end{itemize}\\
 \hline
  \small{\code{profileGLMM\_}\code{Gibbs}} & Draws from the parameters' posterior distribution using a Gibbs sampler. & \begin{itemize}[leftmargin=6pt, itemsep=1pt]
  \item \code{preprocess()} output
  \item Number of posterior draws
  \item Number of burn-in draws
 \end{itemize}\\
 \hline
  \small{\code{profileGLMM\_}\code{postProcess}} & Computes the representative clustering and its characteristics. & \begin{itemize}[leftmargin=6pt, itemsep=1pt]
  \item \code{Gibbs()} posterior samples
 \end{itemize}\\
 \hline
\end{tabular}
\caption{Sequential overview of the main functions in the \pkg{ProfileGLMM} package, where the output of each function provides the mandatory input for the next.}\label{tab:Table1}
\end{table}

The remainder of this section proceeds as follows. We first provide a standard application of the functions using a simulated longitudinal exposome dataset. Subsequently, we demonstrate the method's versatility on a novel example, illustrating how Profile-GLMM can be used to perform piecewise fits.

\subsection{Illustrative longitudinal exposome data}\label{sec:expoLMM}

This section showcases the standard workflow of the package on a simulated dataset, taking inspiration from and simplifying the real dataset utilized in \citet{amestoy_bayesian_2025}. The illustrative dataset consists of a simulated a three-wave longitudinal study of 1500 individuals, totaling 4500 observations, to evaluate the impact of environmental exposures on health.
The dataset contains an indicator variable \texttt{indiv} for each individual, a time varying individual characteristic \texttt{X}, the time of measurement \texttt{t}, two exposure measures \texttt{Exp1,Exp2} and a health outcome \texttt{Y}.

\begin{figure}[ht!]
\includegraphics[width=0.8\linewidth]{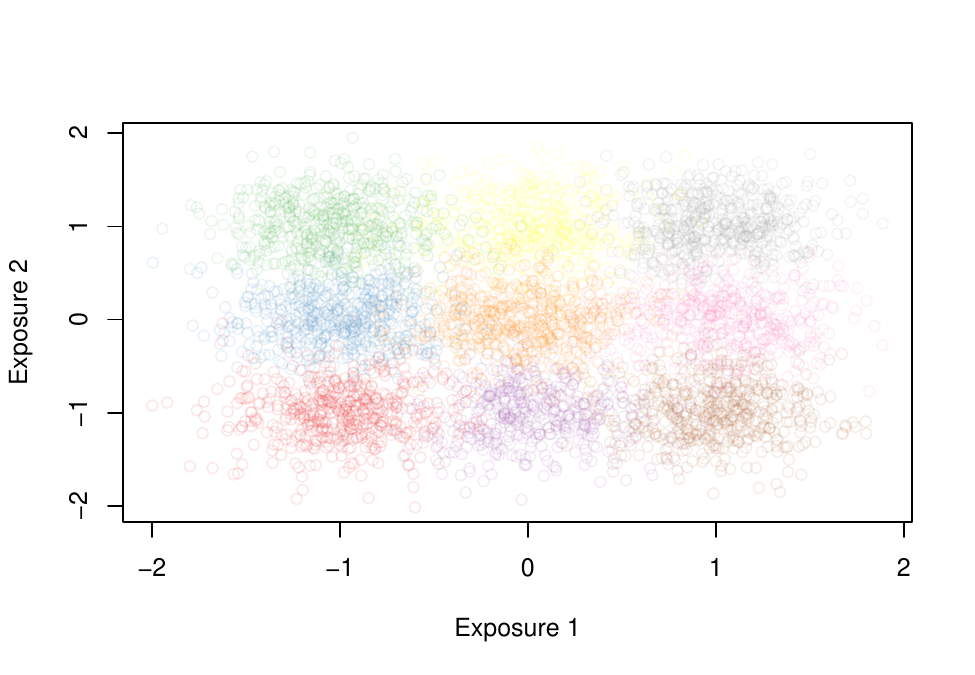} \caption{ Simulated exposure distribution. Color representing the latent cluster.}\label{fig:ExpData}
\end{figure}

\begin{figure}[ht!]
\includegraphics[width=0.95\linewidth]{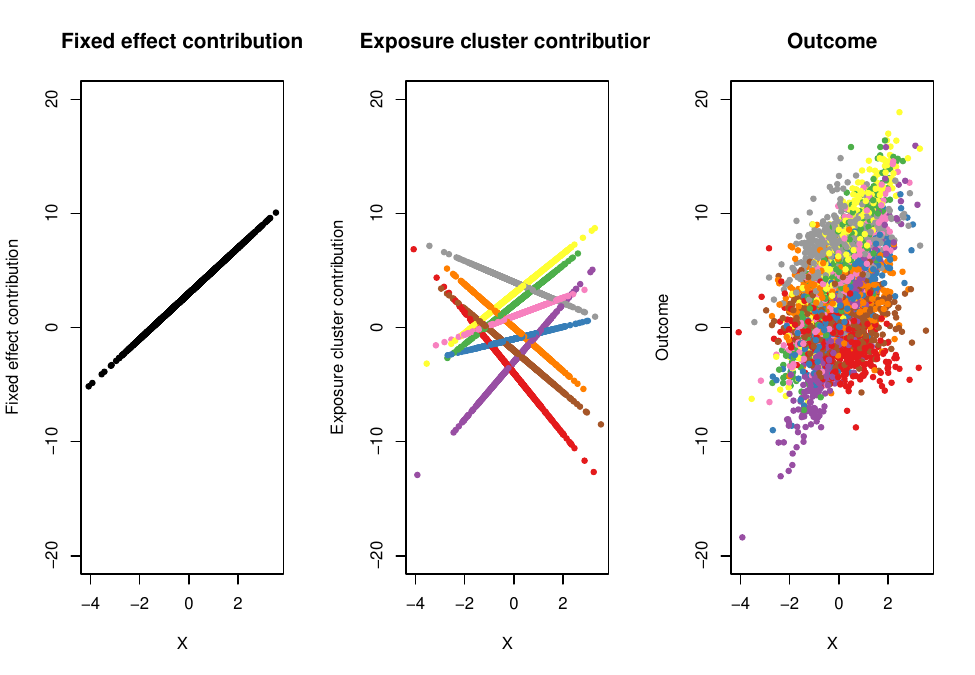} \caption{ Decomposition of the outcome based on the true parameters. Left panel isolates the fixed effect contribution. Central panel shows the effects per latent clusters on the outcome. Right panel is the outcome combining fixed effects, random effects and latent cluster interactions as a function of X.}\label{fig:outcome}
\end{figure}

Figure \ref{fig:ExpData} depicts the distribution of the observed exposures, with colors representing the unknown latent exposure clusters. Each of these nine unique combinations of negative, positive, or neutral exposure values exerts a distinct influence on the health outcome. For observation \(i\) from individual \(j\) (\(\text{indiv}_i=j\)) belonging to latent cluster \(c\), the outcome model is given by:
\[
Y_i = \beta_1 + X_i\beta_2 + t_i\eta_{j}+ \gamma_{c,1}+X_i\gamma_{c,2}+\epsilon_i
\]
Figure \ref{fig:outcome} shows the fixed effect contribution, interaction term per latent cluster and the resulting total outcome \texttt{Y} as a function of \texttt{X}.

The dataset, as well as the underlying parameters that were used to generate it, can be loaded directly from the package using the following R command:

\begin{verbatim}
data("exposure_data")
exp_data = exposure_data$df
theta0 = exposure_data$theta0

head(exp_data)
\end{verbatim}

\begin{verbatim}
#>             X          t indiv        Exp1       Exp2         Y
#> 1 -0.89691455 0.77952297     1 -0.04384600  1.2588400 2.6520873
#> 2  0.18484918 1.70443803     2 -0.66430177 -1.2629737 0.1597814
#> 3  1.58784533 2.22057485     3 -1.40052637 -0.1769742 5.3494451
#> 4 -1.13037567 0.08683717     4 -1.36745207  0.7744430 2.5207198
#> 5 -0.08025176 1.09472696     5  1.19371755  1.4146036 6.6933655
#> 6  0.13242028 2.02427989     6  0.06508781  0.3339544 0.2785757
\end{verbatim}


Once the data is loaded into the R environment, the initial step involves preprocessing the data to prepare all components required for the subsequent posterior sampling procedure. This preparation is achieved using the \texttt{profileGLMM\_preprocess()} function, as shown below.

\begin{verbatim}
covList = {}
covList$FE = c('X')

covList$RE = c('t')
covList$REunit = c('indiv')

covList$Lat = c('X')

covList$Assign$Cont = c('Exp1','Exp2')
covList$Assign$Cat = NULL

covList$Y = c('Y')
dataProfile = profileGLMM_preprocess(regType = 'linear',
                                     covList = covList,
                                     dataframe = exp_data,
                                     nC = 30,
                                     intercept = list(FE = T, RE = F,
                                                        Lat = T))
print(dataProfile)
\end{verbatim}

\begin{verbatim}
#> --- pglmm Data summary ---
#> - Number of observations :  4500 
#> 
#> -- Clustering model summary --
#> - Continuous clustering variables : ' Exp1 Exp2 '
#> 
#> -- Outcome model summary --
#> - Model type:  linear mixed model
#> - Outcome  : ' Y ' 
#> - Fixed effects  : ' Intercept X '
#> - ' indiv ' level random effects  : ' t '
#> - Latent clusters interacting with: ' Intercept X '
\end{verbatim}

\begin{figure}[ht!]
{\centering \includegraphics[width=0.95\linewidth]{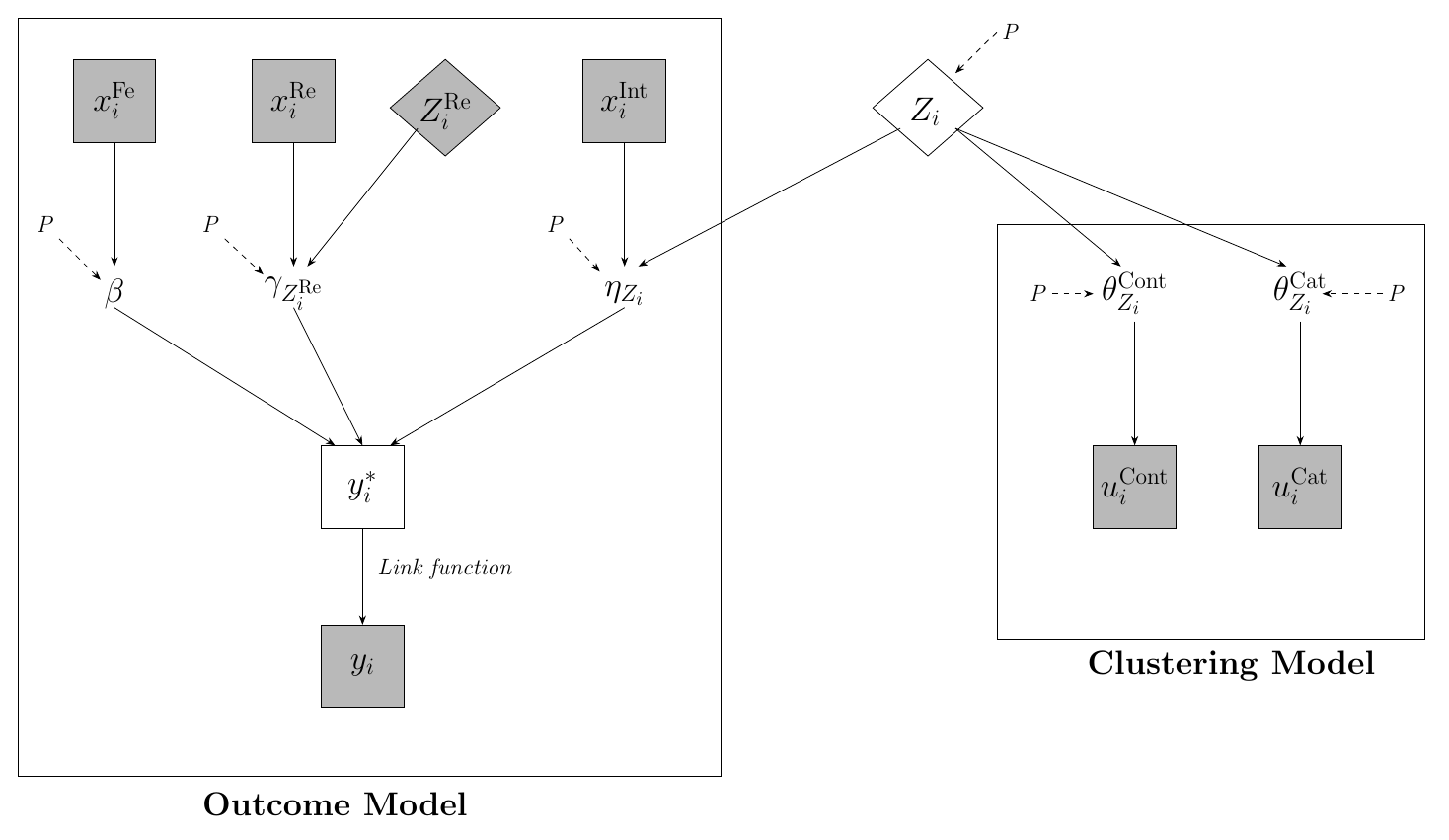} 
}
\caption{Graphical representation illustrating variable dependencies within the ProfileGLMM model. Boxes denote variables, where squares represent general variables (covariates or outcomes) and rhombuses signify cluster membership indicators. Variables with shaded backgrounds are observed (fixed inputs), while unshaded variables are parameters estimated through the MCMC procedure. The configuration of observed variables is managed via the covList entry within the profileGLMM\_preprocess() function call. The influence of prior distributions is indicated by P; and modification procedures are detailed in Section 4. The structure accommodates two primary link functions: the identity function for LMMs (regType = linear) and the Heaviside step function for Probit Mixed Models (regType = probit). }\label{fig:model}
\end{figure}

\sloppy Figure \ref{fig:model} summarizes the general links between the parameters of the \texttt{profileGLMM\_preprocess()} function and the mathematical equations \eqref{eq:LMM}. For our specific example the \texttt{regType} is specified as \texttt{\textquotesingle{}linear\textquotesingle{}}, corresponding to the continuous nature of the outcome variable. The critical input is \texttt{covList}, which defines the structure and covariates for each component of the model:

\begin{itemize}
\item
  The \texttt{FE} field specifies the covariates for the fixed effects (\(x^\text{Fe}\) in Figure \ref{fig:model}, which is \texttt{X} in this case.
\item
  The \texttt{RE} field contains the covariates for the random effect (\(x^\text{Re}\) in Figure \ref{fig:model}, namely \texttt{t}.
\item
  The \texttt{REunit} field indicates the statistical unit for the random effect (\(Z^\text{Re}\) in Figure \ref{fig:model}, which is \texttt{indiv}.
\item
  The \texttt{Lat} field holds the list of covariates that will interact with the latent cluster (\(x^\text{Int}\) in Figure \ref{fig:model}, specified as \texttt{X}.
\item
  The \texttt{Assign} field defines the clustering covariates, subdivided into \texttt{Cont} (continuous) and \texttt{Cat} (categorical) components (\(u^\text{Cont}\) and \(u^\text{Cat}\) in Figure \ref{fig:model}. The two continuous exposure covariates to be clustered, \texttt{Exp1} and \texttt{Exp2}, are indicated here.
\item
  The \texttt{Y} field defines the outcome covariate (\(y\) in figure Figure \ref{fig:model}), which is \texttt{Y}.
\end{itemize}

Finally, the maximal number of clusters considered by the model is set to \texttt{nC\ =\ 30}. The \texttt{intercept} argument specifies that intercepts are included for the fixed effect (\texttt{FE}) and the cluster interaction term (\texttt{Lat}) but excluded from the random effect (\texttt{RE}).

Following data preprocessing, the \texttt{profileGLMM\_Gibbs()} function is invoked to sample from the parameter's posterior distribution. This function requires the preprocessed data object, the total number of Gibbs sampling iterations (\texttt{nIt}), and the count of iterations designated as burn-in (\texttt{nBurnIn}) as inputs. It returns a named list containing the raw Gibbs samples for each parameter. It should be noted that the total length of the resulting thinned MCMC chain will be \texttt{nIt-nBurnIn}.

\begin{verbatim}
MCMC_Obj = profileGLMM_Gibbs(model = dataProfile,
                             nIt = 800,
                             nBurnIn = 200)
\end{verbatim}

\begin{verbatim}
#> Iteration: 0
\end{verbatim}

\begin{verbatim}
print(MCMC_Obj)
\end{verbatim}

\begin{verbatim}
#> --- pglmm_mcmc object containing 600  MCMC samples
\end{verbatim}

\begin{verbatim}
MCMC_ObjExp = MCMC_Obj 
\end{verbatim}

Once samples from the posterior distribution are available, the \texttt{profileGLMM\_postProcess()} function allows for the estimation of population-level parameters, their credible intervals, as well as the characteristics of the representative clustering.

\begin{verbatim}
post_Obj = profileGLMM_postProcess(MCMC_Obj)
\end{verbatim}

\begin{verbatim}
plot(post_Obj,color = colors_for_labels[transfert],
     ylim = c(-2,2),xlim = c(-2,2),
     xlab = 'Exposure 1',
     ylab = 'Exposure 2',pch = 20,
     title = 'Exposure distribution')
points(exp_data$Exp1,
     exp_data$Exp2,
     col = alpha(colors_for_labels[theta0$Lat], 0.1))
\end{verbatim}

\begin{figure}[ht!]
{\centering
\includegraphics[width=0.9\linewidth]{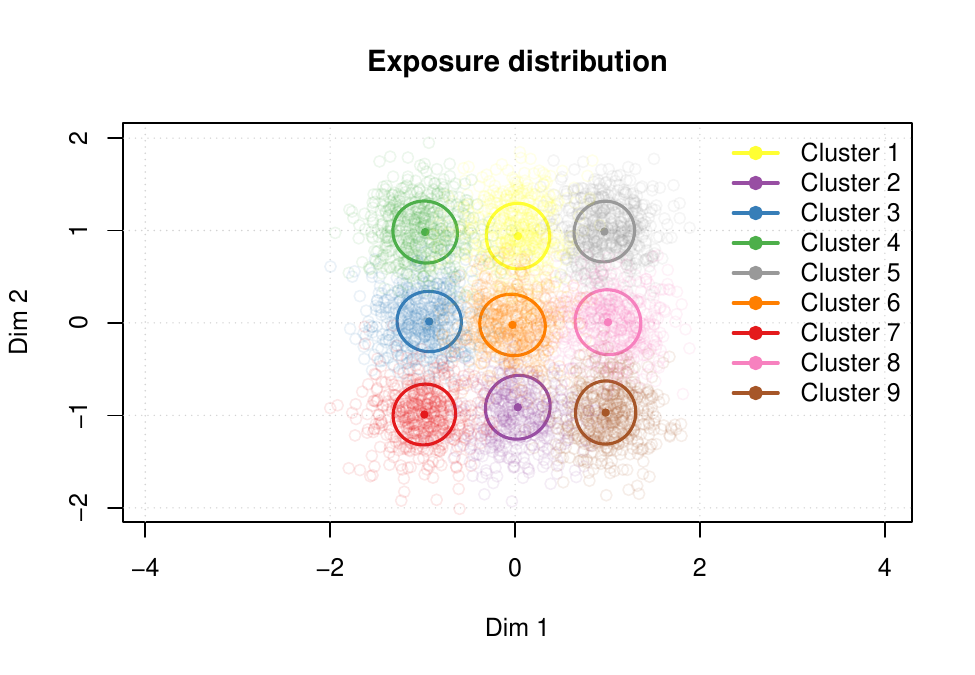} \caption{Profile LMM representative clustering centroids and variance estimates. Points represent observation and their colors match the true latent clusters. Colors of the estimated clusters and true clusters have been matched for readability.}\label{fig:estDist}}
\end{figure}

In this example, the representative clustering is indeed composed of nine clusters, whose centroids closely match the true cluster centroids located on a mesh grid defined by the set \(\{-1, 0, 1\}\). Figure \ref{fig:estDist} illustrates the estimated cluster distribution as well as the estimated cluster membership.

Finally, the package's predictive capability is handled by the \texttt{predict()} method. This function generates predictions for both the latent cluster membership and the outcome for new, out-of-sample observations. Along with the output of \texttt{profileGLMM\_postProcess()}, \texttt{predict()} takes as input the a list of matrices containing the values of \(\mathbf{x}^{\text{FE}}\), \(\mathbf{x}^{\text{Int}}\) and \(\mathbf{u}\) of the new data points. The function returns a list with the following fields: \texttt{\$FE}, the fixed effect contribution to the predicted outcome; \texttt{\$classPred}, the latent cluster's membership prediction; \texttt{\$Int}, the latent cluster's contribution to the predicted outcome; and \texttt{\$Y}, the total predicted outcome, where \texttt{\$FE\ +\ \$Int\ =\ \$Y}.
In the following example, we estimate cluster membership and outcome contributions using the original data:

\begin{verbatim}
pred = predict(post_Obj, dataProfile$d)
\end{verbatim}

\begin{figure}
\includegraphics[width=0.95\linewidth]{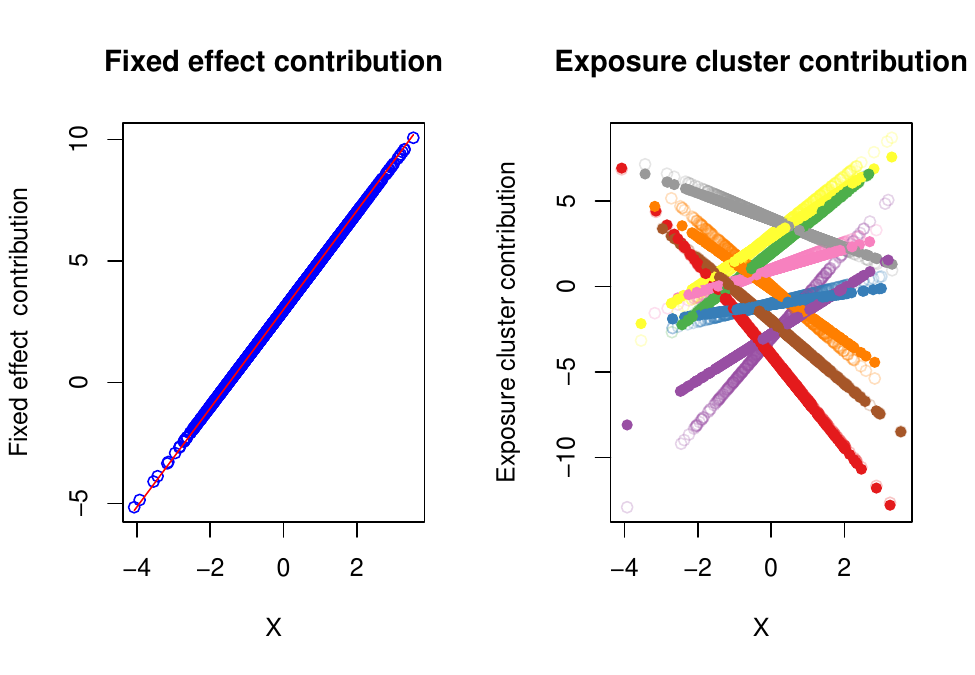} \caption{Left panel isolates the fixed effect contribution with blue being the true contribution red the estimated one. Right panel shows the effects per latent clusters on the outcome with transparent lines being the true contributions.}\label{fig:clusEff}
\end{figure}

From these predictions, the class membership (\texttt{pred\$clasPred}) and its effect can be used to represent each cluster's contribution to the prediction and compared them to the true ones to generate Figure \ref{fig:clusEff}.

\subsection{ProfileGLMM as a piecewise linear fit}\label{sec:piecewise}

In this section, we illustrate a distinctive methodological feature of the Profile-GLMM approach: the direct inclusion of a predictor as a clustering covariate. This capability facilitates the estimation of complex functional relationships, such as a piecewise linear fit, and differentiates the method from current standard Bayesian profile approaches, as outlined in \hyperref[sec:math]{Section 2}. We demonstrate this application using a simple one-dimensional dataset loaded directly from the associated package:

\begin{verbatim}
data("piecewise_data")
piece_data = piecewise_data$df
theta0 = piecewise_data$theta0
head(piece_data)
\end{verbatim}

\begin{verbatim}
#>           x         Y
#> 1  2.001600  3.406760
#> 2  1.350071 10.021829
#> 3  2.851885  5.239479
#> 4 -0.194377 -4.504362
#> 5  1.873669 11.165517
#> 6 -1.765825 -6.674170
\end{verbatim}

\begin{figure}
\includegraphics[width=0.95\linewidth]{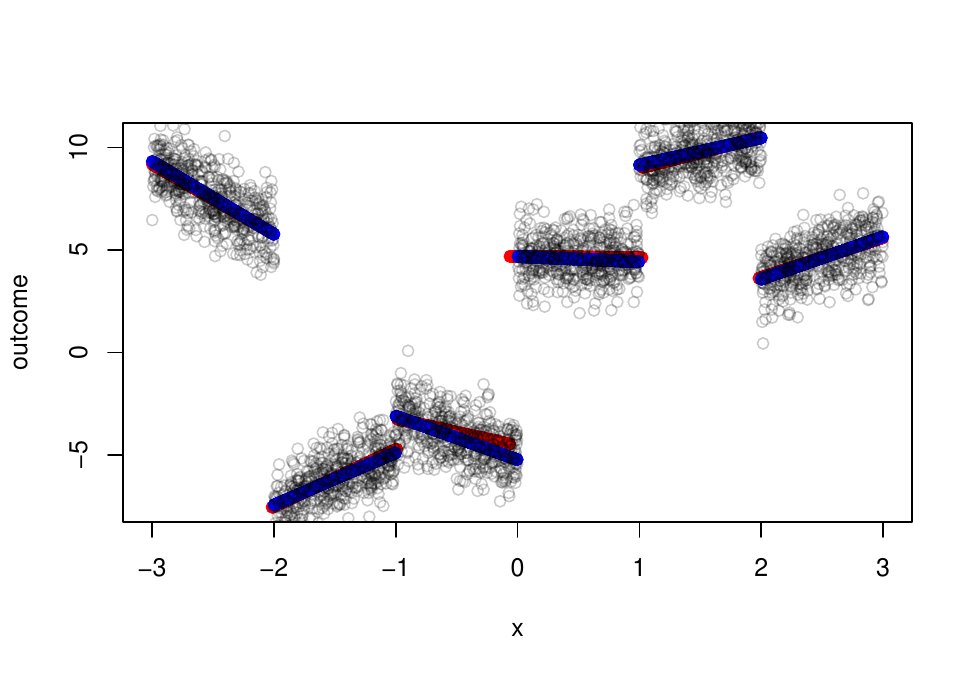} \caption{Blue is the underlying true signal, red is the ProfileGLMM fit}\label{fig:piecewise}
\end{figure}

The simulated data are composed of a single predictor covariate (\(\mathbf{x}\)) and an outcome variable (\(\mathbf{y}\)). The underlying structural relationship between \(\mathbf{y}\) and \(\mathbf{x}\) is defined by a piecewise linear function, perturbed by an additive noise term. This functional decomposition is visually represented in Figure \ref{fig:piecewise}.
The piecewise linear dataset is fitted within the Profile-GLMM framework using the following model specification:

\begin{verbatim}
covList = {}
covList$RE = NULL
covList$FE = NULL
covList$Lat =  c("x")
covList$Assign$Cont = c("x")
covList$Assign$Cat = NULL
covList$REunit = NULL
covList$Y = c('Y')
dataProfile = profileGLMM_preprocess( regType = 'linear',
                                      covList,
                                      piece_data,
                                      20,
                                      intercept = list(FE = T, RE = F,
                                                        Lat = T))
\end{verbatim}

The fixed effect component estimates the overall mean relationship. Crucially, the covariate \(\mathbf{x}\) assumes a dual function: it serves as both the latent clustering covariate and the interaction covariate within the Profile-GLMM structure. This specific modeling configuration enables the Profile-GLMM to perform two simultaneous computational tasks:

\begin{itemize}
\item
  Non-parametrically segment the observations across the domain of \(\mathbf{x}\) based on local heterogeneity (i.e., identifying the piecewise segments or latent regimes).
\item
  Concurrently estimate the cluster-dependent linear trends corresponding to each identified segment.
\end{itemize}

\begin{verbatim}
MCMC_Obj = profileGLMM_Gibbs(dataProfile,
                             nIt = 400,
                             nBurnIn = 200)
post_Obj = profileGLMM_postProcess(MCMC_Obj)
pred = predict(post_Obj, dataProfile$d)
\end{verbatim}

\begin{verbatim}
plot(piece_data$x, pred$Y, col = "red", ylab = 'outcome', xlab='x')
points(piece_data$x, theta0$yLat+theta0$yFe, col = "blue")
points(piece_data$x, piece_data$Y, col = "#00000035")
\end{verbatim}

The resultant reconstruction of the signal is depicted in Figure \ref{fig:piecewise}. The true underlying signal is represented by the blue markers, and the Profile-GLMM estimation is represented by the red markers.

Upon reviewing the results of the Profile-GLMM application, we can clearly observe the method's effectiveness in modeling complex functional relationships. The model successfully identifies six distinct, separate local dynamics within the dataset.
Furthermore, it accurately quantifies the specific impact of each of these identified local dynamics on the overall outcome.

\section{Advanced model configuration and implementation remarks}\label{sec:advanced}

This section contains details of the specification for optional arguments and configurable inputs of all the primary functions. Furthermore, we provide advanced practical remarks pertaining to optimal usage and implementation.

\subsection{\texorpdfstring{\texttt{profileGLMM\_preprocess()}}{profileGLMM\_preprocess()}}\label{sec:prior}

The \texttt{profileGLMM\_preprocess()} function is fundamental, serving as the initialization function that constructs all requisite model components for subsequent functions in the package. This function outputs a \texttt{pglmm\_data} object with a \texttt{print} method associated. Accordingly, we detail two critical configuration aspects. First, we provide an in-depth explanation of the \texttt{covList} entry structure. Second, we describe the procedure for modifying the default prior parameters and initial values for the MCMC chains.

The initial consideration addresses the specification of the model within the \texttt{covList} field. The model specification is designed for flexibility. If specific covariate terms are not utilized in the final model (e.g., no random effects), their corresponding entries within the \texttt{covList} structure may be left empty. This flexibility allows for the complete omission of the random effects component, and the assignment vector \(\mathbf{u}\) may contain either only continuous or only categorical clustering covariates. However, certain components are mandatory for model identifiability. The model must be supplied with at least one clustering covariate as well as components for both fixed effects and the latent cluster interaction.
These mandatory effects can be reduced to only their respective intercepts. Should this be the case, the \texttt{covList\$FE} and \texttt{covList\$Lat} fields can be left empty, provided that the associated intercept indicator for each term is set to \texttt{TRUE}.

We now turn our attention to the modification of prior parameters and the specification of Gibbs sampler initial values. The \texttt{profileGLMM\_preprocess()} function returns an object containing all the necessary components for the Gibbs sampling process. This includes the initialization of parameters for the sampler and the associated hyperparameters for the prior distributions. By default, the prior hyperparameters are configured to the values detailed in Table \ref{tab:priorLMM}, selected primarily to enforce vague and uninformative prior distributions. The initial parameter values for the Gibbs sampler are set by default as random draws from these established prior distributions. The prior's hyperparameters can be readily customized by directly modifying the corresponding fields within the function's output object. The Gibbs sampler initial values are also flexible and can be modified, either by assigning specific values or by initiating a redraw from the prior distribution. The \texttt{prior\_init()} function is provided to facilitate redrawing from the prior, which is particularly useful when the default priors have been customized, requiring an update to the initial parameter settings to reflect these modifications.

\begin{table}[!h]
\centering
\caption{\label{tab:priorLMM}Default settings for the parameters of the outcome model priors.}
\centering
\begin{tabular}[t]{llll}
\toprule
  & Fixed effects & Random effects & Cluster interraction\\
\midrule
Parameters & $(\beta,\sigma^2)$ & $W^{\text{RE}}$ & $W^{\text{Int}}$\\
\midrule
Distributions & $\text{NormGamma}(0,\lambda,a,b)$ & $\mathcal{IW}(\Psi^{\text{RE}},\nu^{\text{RE}})$ & $\mathcal{IW}(\Psi^{\text{Int}},\nu^{\text{Int}})$\\
\midrule
 & $\lambda = 10^{-6}$ & $\Psi^{\text{RE}} = I_{q^{\text{RE}}}$ & $\Psi^{\text{Int}} = I_{q^{\text{Int}}}$\\
Default & $a = 10^{-6}$ & $\nu^{\text{RE}} = q^{\text{RE}}$ & $\nu^{\text{Int}} = q^{\text{Int}}$\\
 & $b = 10^{-6}$ &  & \\
\bottomrule
\end{tabular}
\end{table}

\begin{table}[!h]
\centering
\caption{\label{tab:priorLMMCont}Default settings for the parameters of the clustering model priors Dirichlet process prior.}
\centering
\begin{tabular}[t]{llll}
\toprule
  & Clust cont. & Clust cat. & Dir. proc.\\
\midrule
Parameters & $\theta^{u^{Cont}}_{c}$ & $\theta^{u^{Cont}}_{c,j}$ & $\zeta$\\
\midrule
Distributions & $\text{NIW}(0,\lambda_0,\nu_0,\Phi_0)$ & $\text{Dir}(\rho)$ & $\text{Gamma}(a,b)$\\
\midrule
 & $\lambda_0 = 1$ & $\rho = 1$ & $a = \sqrt{C}$\\
Default & $\Phi_0 = I_{q^{u^{\text{Cont}}}}$ &  & $b = \sqrt{C}$\\
 & $\nu_0 = q^{u^{\text{Cont}}}$ &  & \\
\bottomrule
\end{tabular}
\end{table}

In the following example, we first display the default prior parameters for the fixed effects. Subsequently, we modify the penalty term \(\lambda\) and then utilize \texttt{prior\_init()} to resample the full set of initial values for the Gibbs sampler.

\begin{verbatim}
print(dataProfile$prior$FE$lambda)
\end{verbatim}

\begin{verbatim}
#> [1] 1e-06
\end{verbatim}

\begin{verbatim}
print(dataProfile$theta$beta)
\end{verbatim}

\begin{verbatim}
#> [1] -2981.899
\end{verbatim}

\begin{verbatim}
#Modify the FE prior hyperparameter lambda
dataProfile$prior$FE$lambda = 10
# Update the initialisation of the FE parameter
dataProfile$theta = theta_init(dataProfile$prior,dataProfile$params)
print(dataProfile$theta$beta)
\end{verbatim}

\begin{verbatim}
#> [1] -0.0207891
\end{verbatim}

\subsection{\texorpdfstring{\texttt{profileGLMM\_Gibbs()}}{profileGLMM\_Gibbs()}}\label{profileglmm_gibbs}

\begin{figure}[ht!]
{\centering
\includegraphics[width=0.95\linewidth]{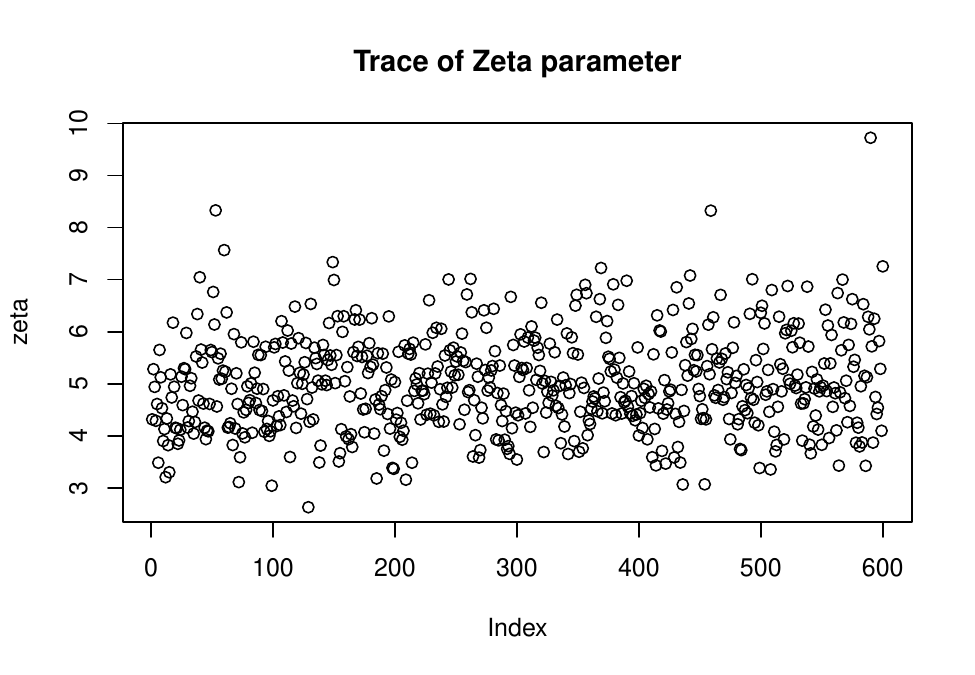} \caption{Trace of the DP concentration parameter after a burn-in of 200 iterations.}\label{fig:zeta}}
\end{figure}

The \texttt{profileGLMM\_Gibbs()} outputs a \texttt{pglmm\_mcmc} object with associated \texttt{print} method.
A critical requirement when utilizing a Markov Chain Monte Carlo (MCMC) sampler is the rigorous verification of chain convergence prior to result analysis. This is typically assessed by visual inspection the trace plot of a population-level parameter, such as \(\zeta\). In our illustrative exposome example, Figure \ref{fig:zeta} displays the trace for \(\zeta\), confirming the adequate convergence of the Gibbs sampler.

It must be noted that the generation of posterior samples is computationally intensive. Our current implementation exhibits competitive efficiency when compared to the established reference package for Bayesian profile regression, \CRANpkg{PReMiuM} \citep{liverani_premium_2015}. A detailed computational time comparison between both packages is provided in Table \ref{tab:pkg-comparison}.

\subsection{\texorpdfstring{\texttt{profileGLMM\_postProcess()}}{profileGLMM\_postProcess()}}\label{profileglmm_postprocess}

Finally, the \texttt{profileGLMM\_postProcess()} returns a \texttt{pglmm\_fit} object with associated \texttt{print}, \texttt{summary}, \texttt{plot} and \texttt{predict} methods. We will start by present the main options of the \texttt{profileGLMM\_postProcess()} function and then briefly present the \texttt{predict()} method.

As discussed in Section \hyperref[postPro]{2.6}, direct use of raw MCMC samples to identify the structure of the latent clusters and their effect on the outcome is difficult. The dedicated function, \texttt{profileGLMM\_postProcess()}, addresses this challenge. This function first computes the cluster co-occurrence matrix from the MCMC partitionings and subsequently derives a single, representative clustering.

Two distinct algorithmic approaches are available to determine the final representative clustering:

\begin{itemize}
\item
  \emph{Least Squares (LS) Method} (\texttt{modeClus=\textquotesingle{}LS\textquotesingle{}}): This approach identifies the sampled cluster assignment that best matches (minimizes the squared error to) the estimated co-occurrence matrix. While this method offers superior computational speed, its robustness is contingent upon the MCMC chain having adequately sampled a partition close to the optimal consensus clustering.
\item
  \emph{Ng's Spectral Clustering} (\texttt{modeClus=\textquotesingle{}NG\textquotesingle{}}): This method employs the spectral clustering algorithm proposed by \citet{ng_spectral_2001} on the similarity matrix. This implementation relies on the \texttt{cluster\_similarity} function provided by the \CRANpkg{Spectrum} package \citep{christopher_r_john_david_watson_spectrum_2019}. Although computationally more demanding than the LS method, the NG approach consistently yields more stable and reliable representative clusterings, leading to its selection as the default configuration.
\end{itemize}

Also note that along with the co-occurrence matrix, the representative clustering estimate and the fixed effect estimates, the \texttt{profileGLMM\_postProcess()} function returns the a posteriori mean estimates for all cluster-specific parameters:

\begin{itemize}
\item
  \texttt{\$clus\$cen}: The center of the Gaussian mixture associated with each identified cluster.
\item
  \texttt{\$clus\$coVar}: The covariance matrix of the Gaussian mixture associated with each identified cluster.
\item
  \texttt{\$clus\$pvec}: The multinomial distribution parameters for each cluster corresponding to each categorical variable in u (omitted in the exposome example).
\item
  \texttt{\$clus\$gamma}: The vector of cluster-specific fixed effects on the outcome variable.
\end{itemize}

\subsection{\texorpdfstring{The \texttt{predict()} method for \texttt{GLMM\_fit} objects}{The predict() method for GLMM\_fit objects}}\label{the-predict-method-for-glmm_fit-objects}

A crucial distinction must be noted between the predictive cluster assignment, \texttt{pred\$classPred} of the \texttt{predict()} function, and the posterior representative clustering indicator of the \texttt{profileGLMM\_postProcess()} function, \texttt{post\_Obj\$Zstar}. The representative clustering indicator, \(\texttt{post\_Obj\$Zstar}\), summarises the draws from the full posterior distribution, relying on both the clustering covariates (\(\mathbf{u}\)) and the observed outcome (\(\mathbf{y}\)). In contrast, \texttt{pred\_Obj\$classPred} represents a purely predictive assignment based exclusively on the clustering covariates \(\mathbf{u}\), as the observed outcome \(\mathbf{y}\) is usually unavailable for new data points.

\subsection{Computational efficiency}\label{computational-efficiency}

Computational efficiency was evaluated by benchmarking the runtimes of the \CRANpkg{ProfileGLMM} package against the established profile regression implementation in the \CRANpkg{PReMiuM} package, with results detailed in Table \ref{tab:pkg-comparison}. The comparison was conducted across various population sizes using a simplified model defined as a basic linear regression (without random effects), incorporating two clustering covariates and a latent-component-specific intercept. The MCMC setting for all trials consisted of 2,000 iterations, with the initial 500 discarded as burn-in. The specifications of the models' comparison code can be found at the end of this section.

We acknowledge that the \CRANpkg{PReMiuM} implementation offers enhanced structural capabilities, such as automated clustering variable selection and unconstrained cluster number allocation, which preclude a direct, strict comparison of execution time. However, the primary objective of this comparison is to demonstrate that the proposed \CRANpkg{ProfileGLMM} package exhibits a computational time complexity of the same order as the current state-of-the-art profile regression package.

\begin{table}[!h]
\centering
\caption{\label{tab:pkg-comparison}Average (over 10 replications) elapsed time for different population size to sample 2000 MCMC draws for the \CRANpkg{ProfileGLMM} and \CRANpkg{PReMiuM} packages.}
\centering
\begin{tabular}[t]{llll}
\toprule
Nb observations & 1 200 & 6000 & 12 000\\
\midrule
ProfileGLMM & 48.02s & 234.53s & 485.47s\\
PReMiuM & 82.83s & 424.56s & 1031.34s\\
\bottomrule
\end{tabular}
\end{table}

The benchmarked snippet of codes are the following:

\begin{verbatim}
    profRegr(yModel = "Normal",
                  xModel = "Normal", nSweeps = 1500, nClusInit = 15,
                  nBurn = 500, data = df, output = "output",
                  covNames = c('Variable1','Variable2'),
                  fixedEffectsNames = c('FixedEffects1','FixedEffects2'),
                  seed = 12345)
    covList = {}
    covList$FE = c('FixedEffects1','FixedEffects2')
    covList$Assign$Cont = c('Variable1','Variable2')
    covList$Assign$Cat = NULL
    covList$Y = c('outcome')
    covList$REunit = NULL
    dataProfile = profileGLMM_preprocess( regtype = 'linear',
                                          covList,
                                          df,
                                          15,
                                          intercept = list(FE = F, 
                                                            RE = F, 
                                                            Lat = T))

    profileGLMM_Gibbs(dataProfile,1500+500+1,500)
\end{verbatim}

\section{Summary}\label{summary}

The \CRANpkg{ProfileGLMM} package successfully integrates the strengths of Bayesian profile regression with the GLMM framework. Incorporating mixed models into the outcome distribution is the central innovation that enables the robust analysis of hierarchical structured data, such as longitudinal measurements, within the unified profile regression paradigm. Furthermore, the inherent flexibility of the method allows practitioners to explore sophisticated modeling approaches, including non-parametric piecewise fits.

In its current release, the package implements mixed models for two primary outcome types: binary data (via a Probit mixed model) and continuous data (via a linear mixed model). The methodology does not impose constraints on the nature of the clustering covariates, accommodating both continuous and categorical variables. Although the underlying inference is based on a Gibbs sampler, the core simulation functions are efficiently implemented in \CRANpkg{Rcpp}. This optimization ensures that \CRANpkg{ProfileGLMM} remains computationally manageable, demonstrating performance that is competitive with established profile regression packages and capable of handling datasets comprising tens of thousands of observations (see \citet{amestoy_bayesian_2025}).

To enhance the practical utility of the package, we provide a suite of dedicated post-processing functions. These tools facilitate the crucial steps of estimating the representative latent cluster partition and quantifying cluster-specific effects. In addition, functions are included for the prediction of the membership of the cluster and the outcome values for new data points. By requiring only minimal mandatory parameters while retaining full flexibility for expert customization, \CRANpkg{ProfileGLMM} is designed to be a practitioner-friendly statistical tool.

We envision several extensions to the current functionalities of \CRANpkg{ProfileGLMM}. Future releases may incorporate a variable selection procedure targeting the clustering covariates, similar to that available in the \CRANpkg{PReMiuM} package, to aid model simplification and remove irrelevant clustering variables. We also plan to expand the range of supported outcomes to include time-to-event data. These methodological improvements will be integrated into forthcoming package releases.
\printbibliography
\end{document}